\documentclass{raa}            
\usepackage{graphicx}
\usepackage{times}
\usepackage{natbib}
\bibpunct{(}{)}{;}{a}{}{,}
\usepackage{amssymb,amsmath}
\usepackage{float}
\usepackage{threeparttable}
\usepackage{booktabs}
\usepackage{adjustbox}
\usepackage{url}            
\usepackage[pagebackref]{hyperref}
\usepackage{ulem}
\usepackage{color}

\usepackage[T1]{fontenc} 

\newcommand{\beq}{\begin{equation}}
\newcommand{\eeq}{\end{equation}}
\newcommand{\bea}{\begin{eqnarray}}
\newcommand{\eea}{\end{eqnarray}}
\newcommand{\bi}{\begin{itemize}}
\newcommand{\ei}{\end{itemize}}
\newcommand{\bfi}{\begin{figure}[!t]
\epsfxsize=9cm
\epsffile}
\newcommand{\bfig}{\begin{figure*}[!t]
\center{}
\epsfxsize=15cm
\epsffile}
\newcommand{\efi}{\end{figure}}
\newcommand{\efig}{\end{figure*}}

\newcommand{\bfx}{{\bf x}}

\begin{document}

  \title{Forecasting the $E_G$ measurements from the photometric and spectroscopic surveys of Chinese Space Station Survey Telescope (CSST)
}

   \volnopage{Vol.0 (2000) No.0, 000--000}      
   \setcounter{page}{1}          

   \author{Yu Song 
      \inst{1}
   \and Yi Zheng
      \inst{1}\thanks{Corresponding author, \url{https://orcid.org/0000-0001-7047-2861}}
}

   \institute{
   School of Physics and Astronomy, Sun Yat-sen University, 
   2 Daxue Road, Tangjia, Zhuhai 519082, China; {\it zhengyi27@mail.sysu.edu.cn}\\
   \and
   CSST Science Center for the Guangdong-Hong Kong-Macau Greater Bay Area, 
   Sun Yat-sen University, Zhuhai 519082, China; \\
\vs\no
   {\small Received 20xx month day; accepted 20xx month day}}

\abstract{We present forecasts for the $E_G$ statistic using redshift distributions of realistic mock galaxy samples from the upcoming Chinese Space Station Survey Telescope (CSST). The dominant uncertainty in $E_G$ stems from the redshift space distortion parameter $\beta$, whose precision limits the overall constraining power. Our analysis shows that CSST will nevertheless achieve $E_G$ constraints at the few-percent level ($3\%-9\%$) over $0 < z < 1.2$, an improvement by a factor of several to an order of magnitude over current observations. Within the $\mu-\Sigma$ modified gravity framework, the parameter $\Sigma_0$, associated with the effective gravitational constant of the Weyl potential, can be constrained to $\sim 5\%$ precision. In a plausible scenario where upcoming spectroscopic surveys determine $\beta$ to $1\%$ accuracy, $E_G$ constraints tighten to the percent level, and $\Sigma_0$ becomes measurable at $\sim 1\%$. For representative modified gravity scenarios, we find that the potential deviations from the  Hu--Sawicki $f(R)$ model and the normal-branch Dvali--Gabadadze--Porrati (nDGP) model remain detectable within the expected sensitivity of CSST. These results demonstrate that CSST will serve as a powerful facility for testing gravity and underscore the essential synergy between photometric weak lensing and spectroscopic surveys in probing cosmic acceleration.
\keywords{gravitational lensing: weak --- (cosmology:) large-scale structure of universe --- cosmology: theory --- cosmology: observations}
}

   \authorrunning{Y. Song \& Y. Zheng }            
   \titlerunning{$E_G$ measurement forecasts for CSST}  

   \maketitle

\section{Introduction} 
\label{sec:Intro}
Since the discovery of the accelerated expansion of the Universe~\citep{1998AJ, 1999ApJ}, understanding its physical origin has remained one of the most profound challenges in modern cosmology. Although some observational tensions persist~\citep{DiValentino2021,Verde2024,Miyatake2023,AbdulKarim2025,DES2026}, the widely accepted \(\Lambda\)CDM model, which incorporates a cosmological constant to drive acceleration, agrees well with a broad range of observations~\citep{WMAP2013,PLANK2018,eBOSS2021,ACT2025,KiDS2025,SPT2025}. However, the theoretical puzzles associated with the cosmological constant, most notably the fine-tuning and coincidence problems~\citep{weinbergreview,Zlatev1999}, motivate the exploration of alternative explanations. These include dynamical dark energy models, which introduce a time-varying component, and modifications to General Relativity (GR) on cosmological scales~\citep{Martin2012,Joyce2015}. Many of these models, however, lead to similar observational signatures, creating degeneracies that complicate their distinction through conventional cosmological probes alone~\citep{Weinberg2013,Joyce2016,Koyama2016}.

In this context, tests of gravity on cosmological scales have become a central focus of next-generation surveys~\citep[e.g.,][]{Alam2021,Frusciante2025}. A particularly powerful probe is the \(E_G\) statistic, first introduced by~\cite{ZhangPJ2007}. This quantity is constructed from a combination of galaxy-galaxy lensing and galaxy clustering measurements. Its key strength lies in its independence from galaxy bias and the amplitude of the matter power spectrum \(\sigma_8\) on linear scales,  which makes \(E_G\) a robust, nearly model-independent diagnostic for deviations from GR, particularly sensitive to modifications in the relationship between the gravitational potentials and the matter distribution.

Current measurements of $E_G$ have relative uncertainties mostly in the range of $10\%-35\%$~\citep{Blake2016,Blake2020,Alam2017,Amon2018,Singh2019,Reyes2010,Jullo2019,Rauhut_2025,Wenzl2025,Wenzl2024,Pullen2016,Zhang2021}, with two outliers reaching $\sim 56\%$ and $78\%$~\citep{de_la_Torre2017}. While these results are generally consistent with GR, they are still limited by statistical power and systematic uncertainties. The next generation of wide-field imaging and spectroscopic surveys, such as  Euclid~\citep{Euclid}, LSST~\citep{LSST2012}, DESI~\citep{DESI2016} and PFS~\citep{PFS12} et al., are designed to deliver order-of-magnitude improvements in data volume and depth. In particular,  China’s upcoming Chinese Space Station Survey Telescope (CSST) will perform a deep, multi-band photometric and slitless spectroscopic survey over approximately \(17{,}500\ \mathrm{deg}^2\)~\citep{CSST2026}. Its unique combination of high-quality imaging and spectroscopic redshift measurements makes it exceptionally well suited for precise \(E_G\) estimation, enabling tomographic gravity tests across a broad redshift range.

In this work, we present forecasts for the constraining power of CSST on the \(E_G\) statistic. We develop a harmonic-space covariance framework that fully accounts for correlations between galaxy-galaxy lensing and galaxy clustering observables, and we employ realistic mock galaxy samples to model the expected redshift distributions and number densities of the CSST photometric and spectroscopic surveys. Our analysis quantifies the statistical precision attainable on \(E_G\) as a function of scale and redshift, identifies the dominant sources of uncertainty, and translates the forecasts into constraints on phenomenological modified gravity parameters within the widely used $\mu-\Sigma$ parametrization~\citep{ZhaoGB2010}.

The paper is organized as follows: Section~\ref{sec:theory} outlines the theoretical and observational definitions of \(E_G\), the error propagation methodology, and the modified gravity parameterization. Section~\ref{sec:CSST_intro} describes the survey specifications of CSST and the mock galaxy samples used in our forecasts. Section~\ref{sec:results} presents the predicted \(E_G\) measurements and their statistical uncertainties, as well as the resulting constraints on modified gravity parameters. Finally, Section~\ref{sec:conclusion} summarizes our conclusions and discusses future prospects.

\section{Analysis framework}
\label{sec:theory}
The $E_G$ statistic combines cosmological weak lensing and galaxy clustering observations to probe the relationships between Bardeen potentials and the matter density fluctuation $\delta$, where Bardeen potentials $\Psi$ and $\Phi$ are potential perturbations in the perturbed Friedmann Robertson Walker (FRW) metric $d\tau^2=(1+2\Psi)dt^2-a^2(1-2\Phi)d\bfx^2$. 

In Fourier space, the theoretical definition of $E_G$ can be expressed as~\citep{ZhangPJ2007}:
 {
\begin{equation}
E_G(k, z) = \frac{c^2 k^2 (\Psi+\Phi)}{3 H_0^2 (1+z) \theta(k)}\,,
\end{equation}
}
where $H_0$ is the Hubble constant, and $\theta(k)$ is the divergence of the peculiar velocity field. On linear scales, $\theta(k) = f(z)\delta(k)$, with $f(z)$ being the linear structure growth rate.

In the context of GR, $\Psi=\Phi$, the theoretical expectation of $E_G$ can be simplified using the Poisson equation, yielding:
\begin{equation}
\label{eq:eg_GR}
E_G(z) = \frac{\Omega_{m,0}}{f(z)}\,,
\end{equation}
where $\Omega_{m,0}$ is the density parameter of the matter at the present day, and $f(z) \approx [\Omega_m(z)]^\gamma$ with $\gamma=0.55$ for GR~\citep{Linder05}. We refer readers to~\cite{Leonard2015} for a systematic study of how variations in both theoretical assumptions and survey parameters can introduce uncertainties into the general relativistic prediction of $E_G$ given in equation~(\ref{eq:eg_GR}), for the $E_G$ measurement in configuration space.

\subsection{Observational definition of \texorpdfstring{$E_G$}{EG}}
\label{subsec:eg_harmonicspace}

The next generation of weak lensing surveys will provide about ten times larger sky coverage than current ones~\citep{laureijs2011euclid,ivezic2019lsst,CSST2026}, which is ideal for harmonic-space statistics.  {This expansion, combined with the growing maturity of accurate $\kappa$ reconstruction methods for masked shear catalogs (e.g.,~\citealt{ShiY2024}), further enhances the practical advantages of power spectrum analysis, including easier theoretical modeling and covariance estimation.}
Consequently, we focus this work on harmonic-space forecasts for $E_G$.
Under the Limber approximation, the angular power spectra of galaxy-galaxy lensing ($C^{g\kappa}_\ell$) and galaxy clustering ($C^{gg}_\ell$) can be written as follows:
\begin{align}
\quad C^{g\kappa}_\ell &= \int dz\, f_{g}(z) W_\kappa(z)\chi^{-2}(z)\, P_{\delta g}\left(k = \frac{\ell + 1/2}{\chi(z)}, z\right), \label{eq:Ckg_GR} \\
C^{gg}_\ell &= \int dz\, f_{g}^2(z)\frac{H(z)}{c} \chi^{-2}(z)\, P_{gg}\left(k = \frac{\ell + 1/2}{\chi(z)}, z\right), \label{eq:Cgg}
\end{align}
where $f_g(z)$ is the galaxy redshift distribution, $\chi(z)$ is the comoving distance to redshift $z$, and $\ell$ is the multipole moment.  {
For a realistic source galaxy sample, the effective lensing kernel is obtained by integrating over the source redshift distribution:
\begin{equation}
W_\kappa(z)
=
\int_z^\infty dz_s \,
\frac{dN_s}{dz_s}
W_\kappa(z,z_s),
\end{equation}
where $dN_s/dz_s$ denotes the normalized redshift distribution of source galaxies, and $W_\kappa(z,z_s)$ is the lensing kernel for a single source plane located at redshift $z_s$, such as
\begin{align}
W_\kappa(z, z_s) &= \frac{3 H_0^2 \Omega_{m,0}}{2 c^2} (1 + z)\, \chi(z) \left[1 - \frac{\chi(z)}{\chi(z_s)}\right], \label{eq:Wkappa} 
\end{align}
where $z_s$ is the source redshift.} The galaxy clustering traces the underlying matter distribution with a scale-independent linear bias on large scales, such that: $P_{\delta g} = b_g\, P_{\delta\delta}, P_{gg} = b_g^2\,P_{\delta\delta}.$

The observational estimator of $E_G(\ell)$ can then be formulated in Harmonic space, in analogy with the CMB lensing definition in~\citep{2016MNRAS.460.4098P}:
\begin{equation}
\hat{E}_G(\ell,z_l) =  {c_\Gamma(\ell) \Gamma( z_l)\,} \frac{\hat{C}^{g\kappa }_\ell}{\beta\, \hat{C}^{gg}_\ell}\,,
\label{eq:eg_l}
\end{equation}
which allows for a scale-dependent and tomographic estimation of $E_G$, suitable for the joint galaxy-galaxy lensing and galaxy clustering analyses. Here, $\hat{C}_\ell^{g\kappa}$ and $\hat{C}_\ell^{gg}$ denote the angular power spectra of galaxy-galaxy lensing and galaxy clustering, respectively. The redshift space distortion (RSD) parameter $\beta$ is defined as $\beta = f(z)/b_g$. On linear scales, $b_g$ and $P_{\delta\delta}$ consequently cancel out in equation~(\ref{eq:eg_l}), rendering the $E_G$ statistic independent of both galaxy bias and the matter power spectrum amplitude~\footnote{For simplicity, we neglect the correction associated with scale-dependent galaxy bias.}. As a result, measurements of (or marginalization over) $b_g$ and $\sigma_8$, which are typically required in other gravity probes, become unnecessary.

In equation~(\ref{eq:eg_l}),
\begin{equation}
  {\Gamma( z_l)} = \frac{\Omega_{m,0}f_g(z_l)H(z_l)}{cW_\kappa(z_l,z_s)} 
\end{equation}
is a scale-independent prefactor that ensures $\hat{E}_G=\Omega_{m,0}/f(z)$ in GR case. in the case of general relativity. Its derivation assumes that the quantities $W_\kappa/f_g$ and $H(z)$ vary slowly over the redshift distribution of the lens galaxies, an approximation that is not exact.

To account for the resulting systematic deviation, we introduce a correction factor $c_\Gamma$, defined as
\begin{equation}
      {c_\Gamma(\ell)} = \frac{cW_\kappa(z_l,z_s)}{f_g(z_l)H(z_l)}\frac{O_\ell^{\delta\delta}}{Q_\ell^{\delta\delta}}\,,
\end{equation}
where
\begin{equation}
O^{\delta\delta}_\ell = \int dz\, f_{g}^2(z)\frac{H(z)}{c} \chi^{-2}(z)\, P_{\delta\delta}\left(k = \frac{\ell + 1/2}{\chi(z)}, z\right)
\end{equation}
and
\begin{equation}
Q^{\delta\delta}_\ell = \int dz\, f_{g}(z) W_\kappa(z)\chi^{-2}(z)P_{\delta\delta}\left(k = \frac{\ell + 1/2}{\chi(z)}, z\right)
\end{equation}
  {are auxiliary quantities entering the correction. In analyses of observational data, these terms are typically evaluated using  GR or MG $P_{\delta\delta}$ calculated from corresponding simulations or non-linear theories (e.g.,~\citealt{Bose2021}) when comparing $E_G$ measurement with GR/MG predictions.
In this work, however, they are computed consistently within the theoretical framework adopted for $C_\ell^{g\kappa}$ and $C_\ell^{gg}$.}

\subsection{Error propagation and covariance estimation}

In this study, we adopt a systematic error propagation framework to evaluate the measurement uncertainties of the $E_G$ statistic. 
This framework is constructed based on the observed ratio of angular power spectra, fully accounting for the statistical correlations between the galaxy-convergence cross-power spectrum and the galaxy auto-power spectrum. It provides a reliable foundation for uncertainty estimation of $E_G$ in cosmological interpretation.

We first define the observed ratio of angular power spectra as:
\begin{equation}
\mathcal{R}(\ell) \equiv \frac{C^{g\kappa}_\ell}{C^{gg}_\ell} \,.
\end{equation}
As both $C^{g\kappa}_\ell$ and $C^{gg}_\ell$ are sourced by the same underlying density field, the cosmic variance is partially canceled in this ratio. Despite this benefit, their strong positive correlation complicates the covariance estimation. As shown later, we explicitly incorporate the full covariance between $C^{g\kappa }_\ell$ and $C^{gg}_\ell$ in our analysis,  to ensure accurate error propagation.

Based on this definition, the $E_G$ statistic in harmonic space can be reexpressed from equation~(\ref{eq:eg_l}) as:
\begin{equation}
E_G(\ell) = \mathcal{R}(\ell)\,\frac{\Gamma}{\beta} \, ,
\end{equation}
and its covariance approximately satisfies:
\begin{equation}
\frac{{\rm Cov}[E_G(\ell)E_G(\ell')]}{E_G(\ell)E_G(\ell')} =
\frac{{\rm Cov}[\mathcal{R}(\ell)\mathcal{R}(\ell')]}{\mathcal{R}(\ell)\mathcal{R}(\ell')} +
\frac{\sigma^2[\beta]}{\beta^2} \, ,
\end{equation}
where $\mathcal{R}(\ell)$ and $\beta$ are assumed to have statistically independent errors. This relation shows that the uncertainty in $E_G$ can be decomposed into two contributions: the measurement error of the power-spectrum ratio and that of the RSD parameter.


Therefore, the construction of the covariance matrix of $\mathcal{R}(\ell)$ is a key component of the uncertainty analysis. 
Based on the theory of multivariate Gaussian random fields, and including possible non-Gaussian correction terms, the full covariance expression is:
\begin{align}
\mathrm{Cov}\big[\mathcal{R}(\ell),\mathcal{R}(\ell')\big]
&= \frac{1}{C_{gg}(\ell)C_{gg}(\ell')} \,
\mathrm{Cov}\!\left[C_{g\kappa}(\ell),C_{g\kappa}(\ell')\right] \nonumber \\[3pt]
&\quad + \frac{C_{g\kappa}(\ell)C_{g\kappa}(\ell')}{C_{gg}^2(\ell) C_{gg}^2(\ell')} \,
\mathrm{Cov}\!\left[C_{gg}(\ell),C_{gg}(\ell')\right] \nonumber \\[3pt]
&\quad - \frac{C_{g\kappa}(\ell')}{C_{gg}^2(\ell')} \frac{1}{C_{gg}(\ell)} 
\mathrm{Cov}\!\left[C_{g\kappa}(\ell),C_{gg}(\ell')\right] \nonumber \\[3pt]
&\quad - \frac{C_{g\kappa}(\ell)}{C_{gg}^2(\ell)} \frac{1}{C_{gg}(\ell')} 
\mathrm{Cov}\!\left[C_{g\kappa}(\ell'),C_{gg}(\ell)\right] .
\end{align}

In practice, the covariance between two angular power spectra can be expanded as:
\begin{equation}
\mathrm{Cov}\left[C_{XY}(\ell), C_{ZW}(\ell')\right] =
\frac{\delta_{\ell\ell'}}{(2\ell+1)\Delta\ell f_{\rm sky}}
\left[ C_{XZ}(\ell)C_{YW}(\ell) + C_{XW}(\ell)C_{YZ}(\ell) \right]
+{\rm Cov^{cNG}}\,,
\end{equation}
where the first term represents the Gaussian cosmic variance contribution, $f_{\mathrm{sky}}$ is the sky coverage fraction, and $\Delta\ell$ is the multipole bin width. The second term denotes the non-Gaussian connected contribution, which becomes significant in the nonlinear regime. 

The Gaussian covariance terms are given by
\begin{align}
\mathrm{Cov}^{(G)}\!\left[C_{g\kappa}(\ell), C_{g\kappa}(\ell)\right]
&= \frac{1}{(2\ell+1)\,\Delta \ell\,f_{\mathrm{sky}}}
\left[
C_{g\kappa}^2(\ell)
+ \left(C_{gg}(\ell) + \frac{1}{n_g}\right)
  \left(C_{\kappa\kappa}(\ell) + \frac{\sigma_\gamma^2}{n_s}\right)
\right], \\
\mathrm{Cov}^{(G)}\!\left[C_{gg}(\ell), C_{gg}(\ell)\right]
&= \frac{1}{(2\ell+1)\,\Delta \ell\,f_{\mathrm{sky}}}
\, 2 \left(C_{gg}(\ell) + \frac{1}{n_g}\right)^2, \\
\mathrm{Cov}^{(G)}\!\left[C_{g\kappa}(\ell), C_{gg}(\ell)\right]
&= \frac{1}{(2\ell+1)\,\Delta \ell\,f_{\mathrm{sky}}}
\, 2 \left(C_{gg}(\ell) + \frac{1}{n_g}\right) C_{g\kappa}(\ell).
\end{align}
where $n_g$ and $n_s$ are the angular number densities of lens and source galaxies, and $\sigma_\gamma$ is the intrinsic shape noise. The covariance is dominated by: (i) cosmic variance, $\propto [(2\ell+1)\Delta\ell f_{\rm sky}]^{-1}$, at large scales -- a contribution that is partially canceled in $E_G$; and (ii) shotnoise ($\propto 1/n_g$) and shape noise ($\propto \sigma_\gamma^2/n_s$) at small scales. The overall signal-to-noise of the $E_G$ estimator is therefore set by the sky fraction $f_{\rm sky}$, the shape noise $\sigma_\gamma$, and the galaxy densities $n_g$ and $n_s$.

The non-Gaussian term arises from the non-Gaussian statistics of the matter density field and can be described by the connected trispectrum:
\begin{align}
&\mathrm{Cov}^{\mathrm{cNG}}\left[C_{XY}(\ell), C_{ZW}(\ell')\right] \nonumber \\
&=\frac{1}{4\pi f_{\mathrm{sky}}} \int \frac{d\chi}{\chi^6}
W_X(\chi)W_Y(\chi)W_Z(\chi)W_W(\chi)
T_m\!\left(\frac{\ell+1/2}{\chi}, \frac{\ell'+1/2}{\chi}, \chi\right),
\end{align}
where $\chi$ is the comoving distance, $W_X(\chi)$ is the weight function of probe $X$, and $T_m(k_1, k_2, \chi)$ is the three-dimensional connected trispectrum of the matter power spectrum, characterizing the four-point correlations beyond the Gaussian approximation. We evaluate the non-Gaussian covariance part via \texttt{pyccl}\footnote{\url{https://github.com/LSSTDESC/CCL}}~\citep{pyccl}.

The full covariance matrix $\mathbf{C} = \mathrm{Cov}[E_G(\ell), E_G(\ell')]$ enables a robust scale-dependent analysis of $E_G(\ell)$ for testing modified gravity theories. To obtain a single, scale-independent constraint, we then combine the measurements across multipole bins using an inverse-variance weighted estimator, yielding
\begin{equation}
\label{eq:EG_noell}
\sigma^2(E_G) = \left( \mathbf{1}^T \, \mathbf{C}^{-1} \, \mathbf{1} \right)^{-1} ,
\end{equation}
where $\mathbf{1}$ is the unit vector with a length equal to the number of $\ell$ bins. 
This estimator corresponds to a minimum-variance combination of the $E_G(\ell)$ measurements, ensuring the tightest possible $E_G$ constraint given the covariance.

The above methodological framework not only provides a solid statistical foundation for the $E_G$ forecast from the CSST survey, but can also be extended to other next-generation cosmological surveys. By fully accounting for the covariance structure among observables, we can robustly assess the ability of the $E_G$ statistic to distinguish GR from modified gravity theories, laying the methodological groundwork for future precision tests of gravity.

\subsection{$\mu$--$\Sigma$ parameterization for modified gravity}
\label{subsec:mu_simga_param}

In modified gravity frameworks, departures from GR can be parametrized through modifications of the two metric potentials $\Psi$ and $\Phi$ using the functions $\mu(a,k)$ and $\Sigma(a,k)$~\citep{ZhaoGB2010}:

\begin{equation}
    k^2 \Psi 
    = -4\pi G a^2 \left[1+\mu(k,z)\right] \rho\,\delta ,
\end{equation}
\begin{equation}
    k^2 \left(\Psi + \Phi\right)
    = -8\pi G a^2 \left[1+\Sigma(k,z)\right] \rho\,\delta .
\end{equation}
Here, $\mu$ modifies the Newtonian potential responsible for structure growth, while $\Sigma$ determines the amplitude of the Weyl potential directly probed by lensing.

Under this parametrization, the theoretical prediction of the $E_G$ statistic can be written as :
\begin{equation}
    E_G(k,z) = 
    \frac{\Omega_{m,0}\, \left[1+\Sigma(k,z)\right]}
         {2\, f(k,z)} \,.
    \label{eq:EG_muSigma}
\end{equation}  
In modified gravity scenarios, the linear growth rate $f(k,z)$ is computed by integrating the modified growth equation with the effective gravitational strength determined by $\mu(k,z)$~\citep{pyccl}. 

For simplicity, in this work we will ignore the scale dependence  and adopt phenomenological late-time forms for $\mu$ and $\Sigma$:  
\begin{equation}
    \mu(z) = \mu_0\, \frac{\Omega_\Lambda(z)}{\Omega_\Lambda(0)}\,, 
    \qquad 
    \Sigma(z) = \Sigma_0\,\frac{\Omega_\Lambda(z)}{\Omega_\Lambda(0)}\,,
\end{equation}
which preserve early-universe constraints but allow deviations from GR at low redshift~\citep[e.g.,][]{Simpson2013,Ade2016}. 

By substituting the solution of the growth equation into equation~(\ref{eq:EG_muSigma}), one obtains the theoretical prediction $E_G^{\mathrm{th}}(z;\mu_0,\Sigma_0)$, which can be directly compared to observations. By assuming the $E_G$ measurements of different $z$ are uncorrelated, the Gaussian likelihood is given by
\begin{equation}
    -\ln\mathcal{L} \propto \chi^2(\mu_0,\Sigma_0)
    =
    \sum_i
    \frac{
    \left[
        E_G^{\mathrm{obs}}(z_i) -
        E_G^{\mathrm{th}}(z_i;\mu_0,\Sigma_0)
    \right]^2}
    {\sigma_{E_G(z_i)}^2} .
\end{equation}
The best-fit values and confidence intervals for $(\mu_0,\Sigma_0)$ can be obtained by minimizing $\chi^2$.

\subsection{Representative modified gravity models}

In addition to the $\mu$--$\Sigma$ parameterization, we consider representative modified gravity models that introduce scale-dependent deviations from general relativity (GR). 

The $f(R)$ gravity model extends the Einstein--Hilbert action by replacing the Ricci scalar $R$ with a general function $R + f(R)$, introducing an additional scalar degree of freedom that mediates a fifth force~\citep{Hu2007,2010Felice}. This modification leads to an enhancement of structure growth below a characteristic Compton wavelength, resulting in a scale-dependent gravitational interaction. In this work, we adopt the commonly used Hu--Sawicki form of $f(R)$ gravity, characterized by the present-day scalar field amplitude $|f_{R0}| = 10^{-5}$, which determines the strength of the deviation from GR~\citep{2024Kou,Thomas2015a}.

We also consider the normal branch of the Dvali--Gabadadze--Porrati (nDGP) model, which arises from a braneworld scenario in which gravity leaks into extra dimensions on large scales~\citep{Dvali2000}. The model is characterized by the crossover scale $r_c$, beyond which gravitational interactions are modified~\citep{2022Sakr}. We adopt two benchmark values for the crossover scale, $r_c H_0 = 1$ and $r_c H_0 = 5$, corresponding to moderate and weak departures from GR, respectively, with the latter being closer to the GR limit. In this regime, the modifications are approximately scale-independent on linear scales, while still producing observable effects in large-scale structure.

These models provide complementary examples of modified gravity, with $f(R)$ exhibiting strong scale dependence and nDGP primarily affecting the amplitude of gravitational interactions, enabling a systematic assessment of the sensitivity of $E_G$ to different classes of deviations from GR.


\section{CSST surveys}
\label{sec:CSST_intro}
\begin{figure}[t!]
        \includegraphics[width=0.48\textwidth]{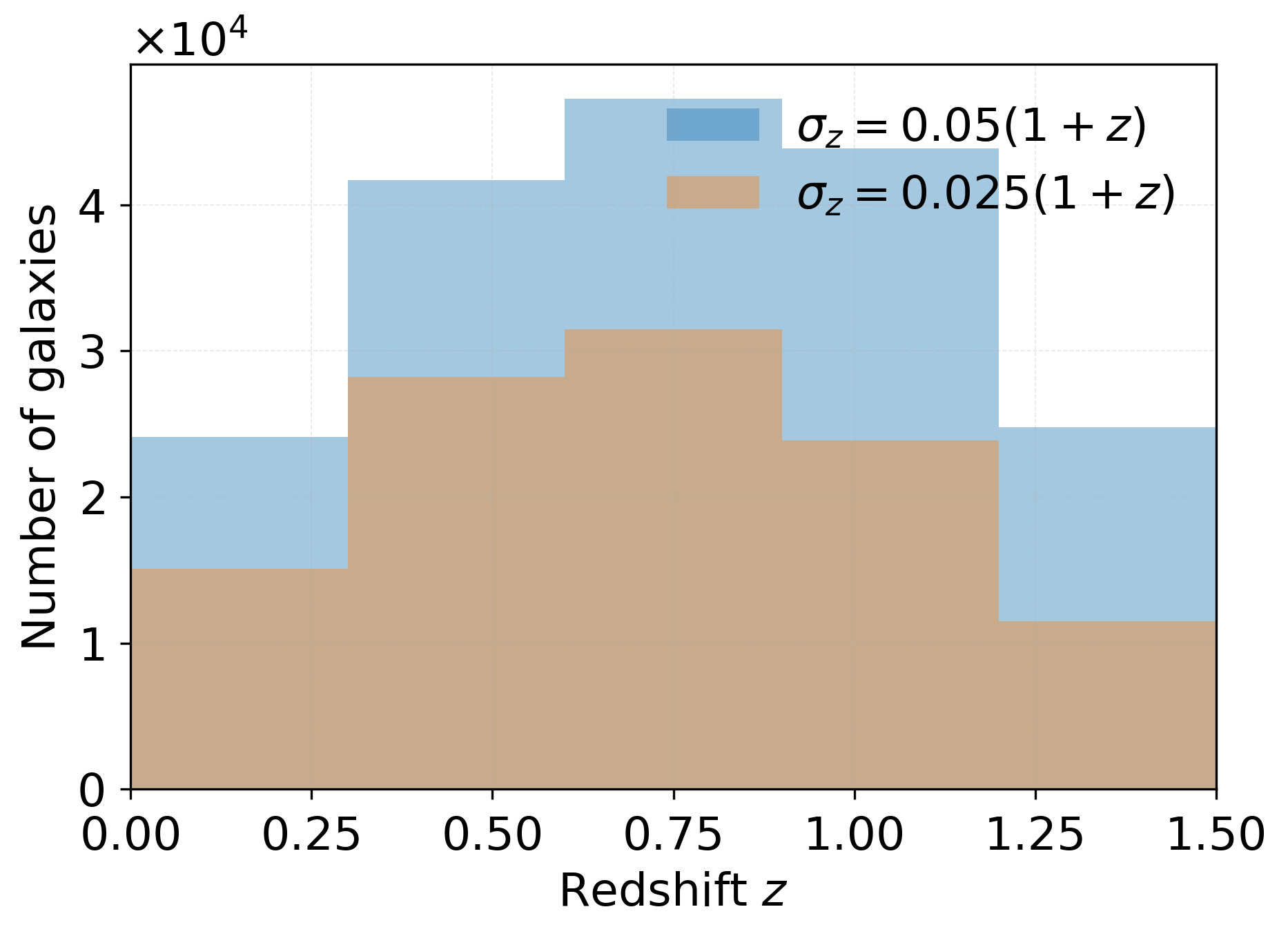}
        \includegraphics[width=0.48\textwidth]{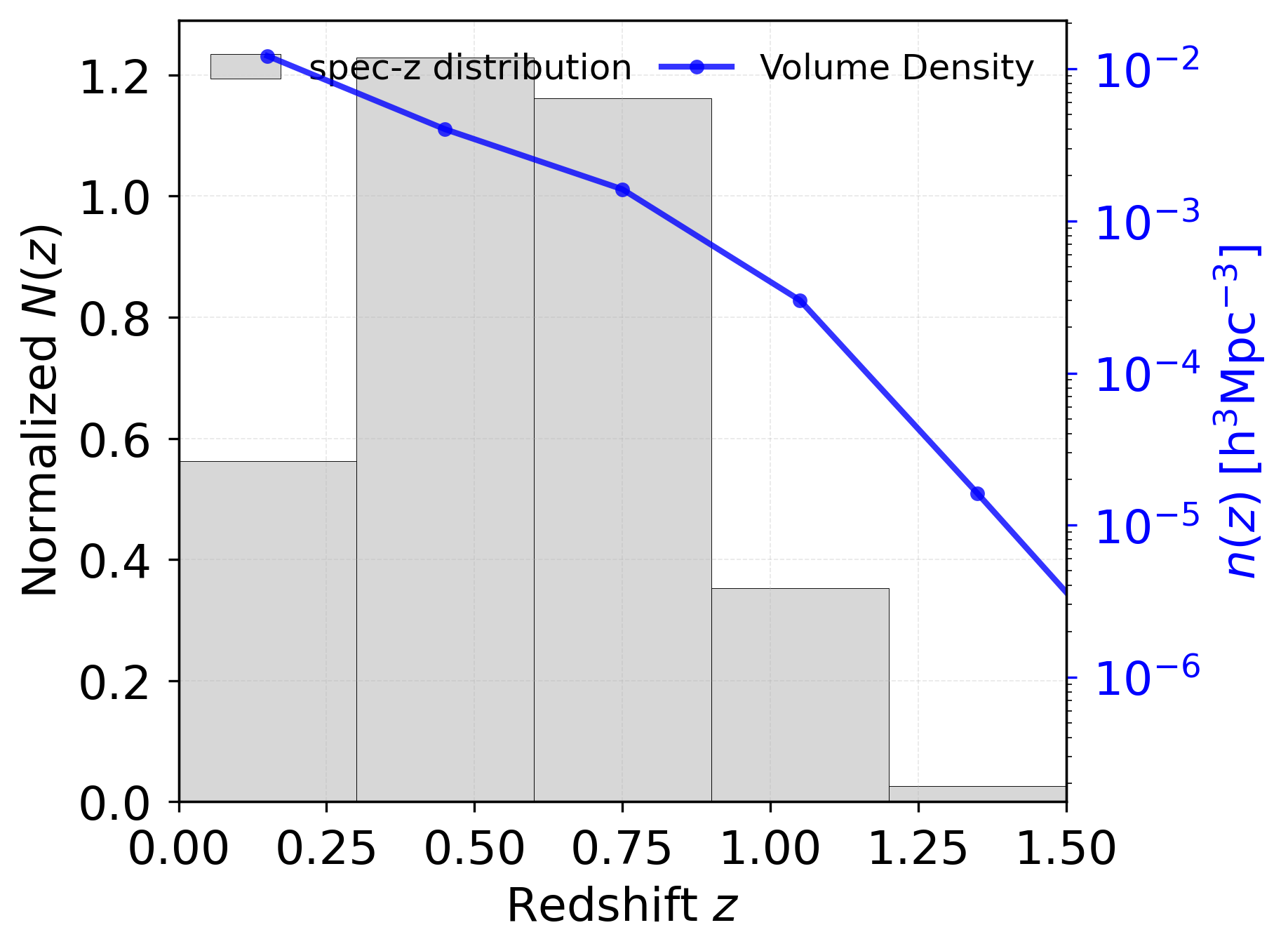}
        \caption{\textit{Left:} Redshift distributions of galaxies in the CSST photometric survey. 
        The distributions are derived from a sub-set of the COSMOS catalog within $1.7\,\mathrm{deg}^2$. 
        We adopt the $\sigma_z = 0.025(1+z)$ sample in our analysis. \textit{Right:} Galaxy redshift distribution for the CSST slitless spectroscopic (spec-$z$) survey. 
        The distribution is normalized following ~\citep{2019ApJ...883..203G}, and the comoving number density evolution 
        with redshift is also shown.}
        \label{fig:gal_dist}
\end{figure}
\begin{table}[t!]
\centering

\begin{tabular}{ccccc}
\hline
Redshift range & $V_{\mathrm{survey}}$ [$h^{-3}\,\mathrm{Gpc}^3$] & $\bar{n}_g$ [$10^{-4}\,h^3\,\mathrm{Mpc}^{-3}$] & $b_g(z)$ & $\sigma(\beta)/\beta$ \\
\hline
$0.0 < z < 0.3$ & 1.03  & 121.93 (1283.6) & 1.126 & 8.73\% \\
$0.3 < z < 0.6$ & 5.41  & 40.18  (456.5)  & 1.378 & 4.02\%\\
$0.6 < z < 0.9$ & 10.51 & 16.18  (262.0)  & 1.63 & 3.14\%\\
$0.9 < z < 1.2$ & 14.64 & 3.01   (142.8)  & 1.882 &3.28\% \\
$1.2 < z < 1.5$ & 17.48 & 0.16   (57.64)  & 2.134 & -\\
\hline
\end{tabular}
\caption{Statistical properties of the CSST spec-$z$ (photo-$z$) galaxy surveys in five redshift bins over a sky coverage of $17{,}500~\mathrm{deg}^{2}$.  Values in parentheses refer to the photo-$z$ sample. The galaxy bias is modeled as $b_g(z) = 1 + 0.84z$ for both surveys~\citep{2024MNRAS.527.3728D}. The error prediction of the $\beta$ parameter in the lens redshift bin comes from~\cite{LiZY2026}. The intrinsic shape noise $\sigma_\gamma$ is set to be 0.2~\citep{LinHJ2022}.}
\label{tab:redshift_bins}
\end{table}


CSST is a 2 meter space telescope for the Stage IV dark energy experiment. It will operate in the same orbit as the China Space Station and is designed to perform a high resolution, wide field photometric and spectroscopic survey covering approximately 17,500 deg$^2$ of the sky over a 10-year mission duration~\citep{CSST2026}.

The CSST main focal plane is divided into 7 photometric imaging bands (NUV, $u$, $g$, $r$, $i$, $z$, and $y$) and 3 slitless spectroscopic bands (GU, GV, and GI), covering the wavelength range of 255--1000 nm with an average spectral resolution of $R \geq 200$. The 17,500 deg$^2$ slitless spectroscopic wide-field survey is performed simultaneously with the multi-color imaging survey, achieving magnitude limits of not less than 22 mag for the GU band and not less than 23 mag for both the GV and GI bands.~\citep{CSST2026}

Due to the large and overlapping sky coverage of the CSST photometric and spectroscopic surveys, it will be possible to jointly probe the growth and geometry of the Universe using multiple cosmological observables. These include weak gravitational lensing~\citep{2019ApJ...883..203G,LinHJ2022,XiongQ2025}, cluster number counts~\citep{ZhangYF2023,ChenMJ2025}, the Alcock Paczynski (AP) effect~\citep{XiaoL2023}, RSD effect~\citep{2019ApJ...883..203G,MiaoHT2023}, baryon acoustic oscillations (BAO)~\citep{DingZJ2024,MiaoHT2024}, among others. The CSST is therefore expected to serve as a powerful instrument for investigating both the expansion history of the Universe and the growth of cosmic structures. In particular, by combining various cosmological probes within the CSST survey, one can obtain more robust and precise constraints on fundamental cosmological parameters, effectively breaking degeneracies that arise when using individual probes alone~\citep{2019ApJ...883..203G,CSST2026}. 

Similar to~\cite{Ding2024}, in this work we make use of redshift distribution of the mock galaxy catalogs for the CSST surveys provided by ~\cite{Cao2018,2019ApJ...883..203G}, 
which are based on the COSMOS galaxy catalog~\citep{Capak2007,Ilbert2009} and the $z$COSMOS spectroscopic samples~\citep{Lilly2007,Lilly2009}. 
~\citep{Cao2018} constructed two photometric-redshift (photo-$z$, $z_p$) samples using a sub-set (95\% and  58\%) of the COSMOS catalog, with 
photometric redshift uncertainties respectively being 
$\sigma_{z_p} = 0.025(1+z_p)$ and $\sigma_{z_p} = 0.05(1+z_p)$. 
The left panel of figure~\ref{fig:gal_dist} shows the redshift distributions of galaxies in the range $0<z<1.5$ for these two photo-$z$ samples, and in our analysis, we adopt the photo-$z$ sample with $\sigma_{z_p} = 0.025(1+z_p)$ as the fiducial mock configuration.



We also consider the galaxy redshift distribution of the CSST slitless spectroscopic (spec-$z$, $z_s$) survey. Following~\cite{Ding2024},  we extract from zCOSMOS a sub-sample with robust redshift measurements, representing about 80 per cent of the full sample and covering $0<z\le1.5$. Its normalized spectroscopic redshift distribution (right panel of figure~\ref{fig:gal_dist}) exhibits a steep decline beyond $z=1.0$. Moreover, we account for redshift uncertainties arising from the limited spectroscopic resolution of CSST slitless spectroscopy. Our baseline model incorporates the error parameterization $\sigma_{z_s}=0.002(1+z_s)$~\citep{GongY2019} and a redshift success rate of the form $f^z_{\rm eff}=f^0_{\rm eff}(1+z_s)$~\citep{WangY2010}, where $f^0_{\rm eff}$ denotes the success fraction at $z=0$. We adopt $f^0_{\rm eff}=0.5$  as our fiducial choice and show the corresponding comoving volume number density of spec-$z$ galaxy as the solid line in the right panel of figure~\ref{fig:gal_dist}.
 
These photo-$z$ and spec-$z$ mock samples provide a consistent framework to evaluate the CSST survey performance for weak lensing and galaxy clustering analyses. The redshift binning scheme, comoving survey volume, galaxy number density, galaxy bias and the corresponding error prediction of $\beta$ for each redshift bin are summarized in Table~\ref{tab:redshift_bins}.

\section{Results}
\label{sec:results}

\subsection{$E_G$ predictions}
\label{subsec:result_EG}
\begin{figure}[t!]
    \centering
    \includegraphics[width=1\textwidth]{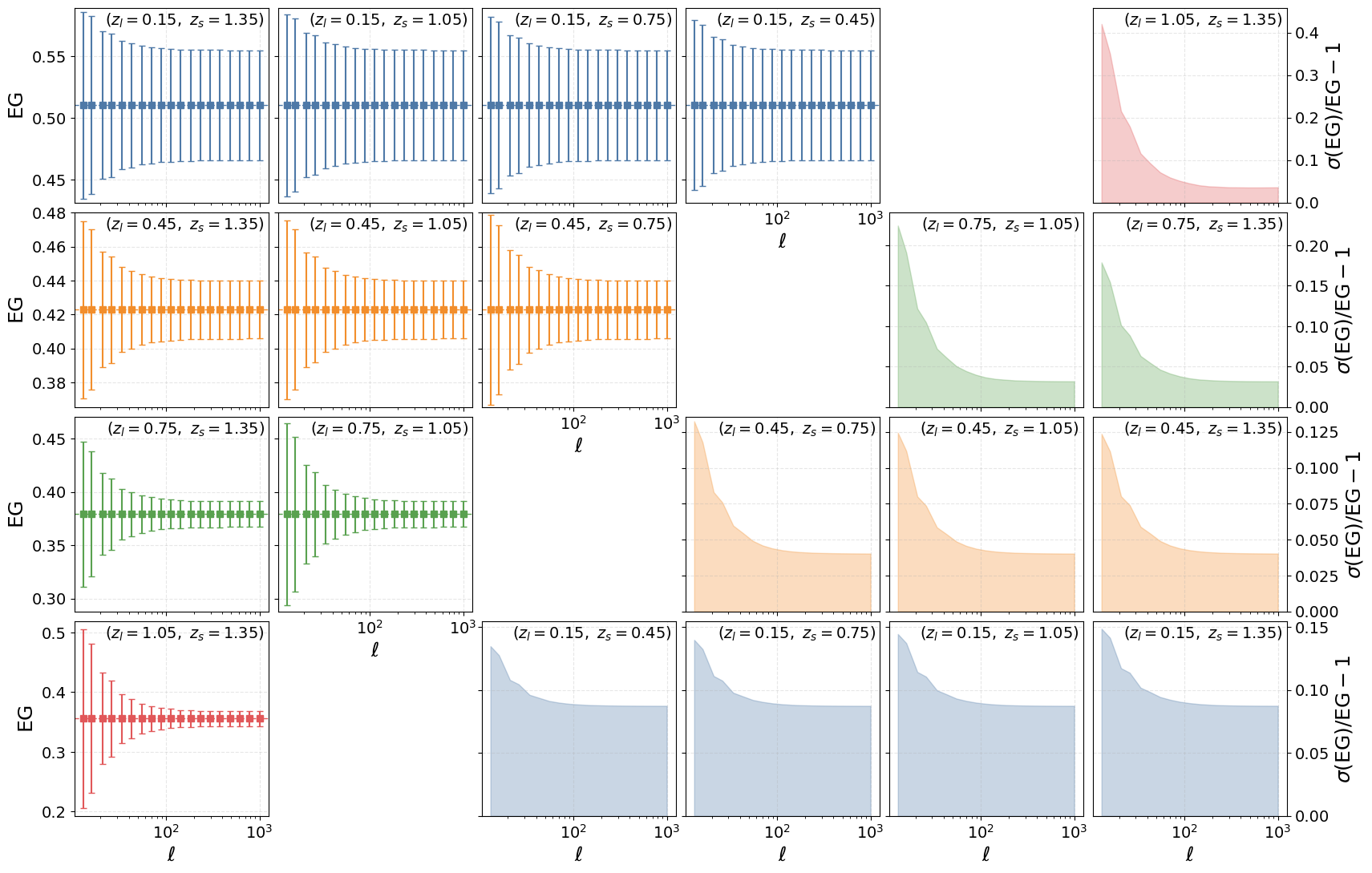}
    \caption{Forecasted $E_G$ measurements in harmonic space. 
    The upper panel shows $E_G(\ell)$ for different combinations of lens and source bins, 
    with error bars denoting Gaussian uncertainties. 
    The lower panel presents the relative statistical uncertainty, 
    $\sigma(E_G)/E_G - 1$, as a function of multipole $\ell$.
    For clarity, each sub-panel is labeled by the mean redshifts $\bar{z}_l$ and $\bar{z}_s$ of the lens and source samples, respectively.
    The corresponding redshift ranges are defined in Table~\ref{tab:redshift_bins}.
    }
    \label{fig:eg_l}
\end{figure}
\begin{figure}[t!]
    \centering
    \includegraphics[width=0.49\textwidth]{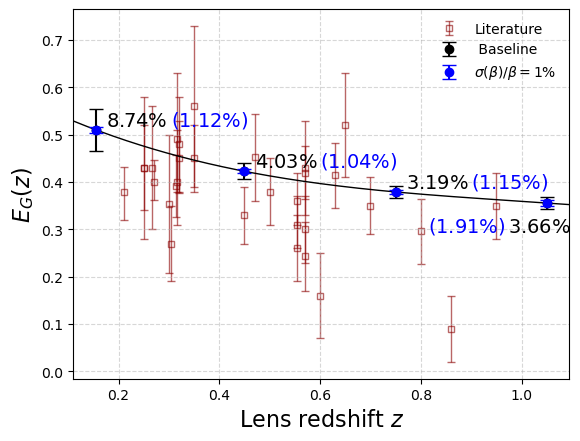}
    \includegraphics[width=0.49\textwidth]{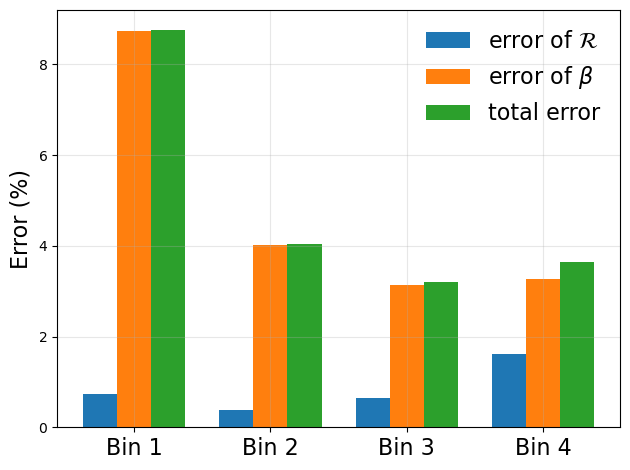}
    \caption{\textit{Left:} redshift evolution of the $E_G$ statistic after combining all source bins for each lens bin. Blue points with error bars show the results assuming $\sigma(\beta)/\beta = 1\%$. Percentile numbers beside data points represent the fractional errors of $E_G$. Data represented by hollow red squares show current $E_G$ measurements from the literature collected in table~\ref{tab:eg_literature}. For clarity and compactness of the figure, the data point at $z=1.5$ in table~\ref{tab:eg_literature} ($E_G = 0.30 \pm 0.05$~\citep{Zhang2021}) is not shown. \textit{Right:} Fractional error contributions to the $E_G$ estimator. The bars show the total statistical uncertainty, the error from the measurement of the structure growth rate $\beta$, and the combined contributions from all other sources.}
    \label{fig:eg_sum}
\end{figure}

\begin{table}[t!]
	\centering
	\renewcommand{\arraystretch}{1.25}
	\begin{tabular}{@{\hspace{4mm}}c@{\hspace{6mm}}c@{\hspace{6mm}}l@{\hspace{10mm}}c@{\hspace{6mm}}c@{\hspace{6mm}}l@{\hspace{4mm}}}
		\hline
        $z$ & $E_G$ & ref  & $z$ & $E_G$ & ref \\
        \hline
        
        0.21  & $0.38^{+0.052}_{-0.061}$ & ~\citep{Singh2019}
        & 0.50  & $0.38 \pm 0.07$          & ~\citep{Rauhut_2025} \\
        
        0.25  & $0.43 \pm 0.09$          & ~\citep{Blake2020}
        & 0.554 & $0.26 \pm 0.07$          & ~\citep{Amon2018} \\
        
        0.25  & $0.43 \pm 0.15$          & ~\citep{Rauhut_2025}
        & 0.555 & $0.36^{+0.06}_{-0.05}$   & ~\citep{Wenzl2024} \\
        
        0.267 & $0.43 \pm 0.13$          & ~\citep{Amon2018}
        & 0.555 & $0.31^{+0.06}_{-0.05}$   & ~\citep{Wenzl2025} \\
        
        0.27  & $0.40^{+0.046}_{-0.038}$ & ~\citep{Singh2019}
        & 0.57  & $0.30 \pm 0.07$          & ~\citep{Blake2016} \\
        
        0.30  & $0.354 \pm 0.146$        & ~\citep{Li2025}
        & 0.57  & $0.42 \pm 0.056$         & ~\citep{Alam2017} \\
        
        0.305 & $0.27 \pm 0.08$          & ~\citep{Amon2018}
        & 0.57  & $0.43 \pm 0.10$          & ~\citep{Jullo2019} \\
        
        0.315 & $0.392 \pm 0.065$        & ~\citep{Reyes2010}
        & 0.57  & $0.243 \pm 0.073$        & ~\citep{Pullen2016} \\
        
        0.316 & $0.40^{+0.11}_{-0.09}$   & ~\citep{Wenzl2024}
        & 0.60  & $0.16 \pm 0.09$          & ~\citep{de_la_Torre2017} \\
        
        0.316 & $0.49^{+0.14}_{-0.11}$   & ~\citep{Wenzl2025}
        & 0.63  & $0.414 \pm 0.069$        & ~\citep{Li2025} \\
        
        0.32  & $0.48 \pm 0.10$          & ~\citep{Blake2016}
        & 0.65  & $0.52 \pm 0.11$          & ~\citep{Rauhut_2025} \\
        
        0.32  & $0.45^{+0.080}_{-0.073}$ & ~\citep{Singh2019}
        & 0.70  & $0.35 \pm 0.06$          & ~\citep{Rauhut_2025} \\
        
        0.35  & $0.56 \pm 0.17$          & ~\citep{Rauhut_2025}
        & 0.80  & $0.296 \pm 0.069$        & ~\citep{Li2025} \\
        
        0.35  & $0.45 \pm 0.07$          & ~\citep{Blake2020}
        & 0.86  & $0.09 \pm 0.07$          & ~\citep{de_la_Torre2017} \\
        
        0.45  & $0.33 \pm 0.06$          & ~\citep{Blake2020}
        & 0.95  & $0.35 \pm 0.07$          & ~\citep{Rauhut_2025} \\
        
        0.47  & $0.452 \pm 0.092$        & ~\citep{Li2025}
        & 1.50  & $0.30 \pm 0.05$          & ~\citep{Zhang2021} \\
        
        \hline
	\end{tabular}
\caption{Compilation of $E_G$ measurements from the literature, ordered by redshift.}
\label{tab:eg_literature}
\end{table}

We present predictions for the $E_G$ statistic based on the mock CSST redshift distributions detailed in Section~\ref{sec:CSST_intro}. Figure~\ref{fig:eg_l} summarizes the results. The top left panels show $E_G(\ell)$ for various combinations of lens and source redshift bins, while the lower right panels display the relative statistical uncertainty, $\sigma(E_G)/E_G$, as a function of multipole $\ell$ across 20 log-spaced bins between $\ell=10$ and $\ell=1000$.

Our forecasts indicate that few-percent level precision on $E_G$ is achievable over a wide range of angular scales and lens-source redshift pairs. The constraints exhibit a distinct scale dependence: at large scales ($\ell < 100$), cosmic variance dominates, leading to larger uncertainties. At smaller scales ($\ell > 100$), the precision improves significantly, with constraints reaching $\sim 4\%$ across various lens redshift bins.   {This scale-dependent behavior can be understood by checking the $\beta$ constraints~\citep{LiZY2026} shown in the last column of Table~\ref{tab:redshift_bins}.} At small scales, the $E_G$ constraints are predominantly limited by the precision on $\beta$, which explains the flattening of the $E_G$ uncertainty curves in this regime.

To obtain a scale-independent $E_G$ constraint, we combine the $E_G(\ell)$ measurements across different multipoles using equation~(\ref{eq:EG_noell}). In this process, we adopt the simplifying assumption that, for a given lens redshift bin, we treat all available source redshift bins as a single combined source sample to maximize the lensing signal. The resulting integrated $E_G$ constraints as a function of the lens redshift are shown in the left panel of figure~\ref{fig:eg_sum}. The predicted $E_G$ measurements can reach $3\%-9\%$ over the lens redshift range $0<z<1.2$ --- a factor of several to an order of magnitude improvement over current observations. These results demonstrate that CSST will be a highly competitive facility for future $E_G$ studies across a broad range of scales and redshifts~\citep{Pullen2015,Pourtsidou2016,Abidi2023,Patki2025,Leonard2025,Li2025}.

As anticipated, the total uncertainty on $E_G$ is dominated by the limited precision in determining $\beta$ through the RSD effect. This is explicitly demonstrated in the right panel of figure~\ref{fig:eg_sum}, which compares the fractional error contributions from the ratio $\mathcal{R} = C^{g\kappa}/C^{gg}$ and from $\beta$. Consequently, improving the overall $E_G$ precision hinges primarily on obtaining tighter constraints on $\beta$. This conclusion holds true for the current $E_G$ detections as well, and is supported by the findings of~\citep{Grimm2024}, who achieved $6-11.3\%$ precision of $E_G$ measurements by obtaining improved $\beta$ constraints through a model-independent Gaussian process reconstruction that combined multiple literature measurements. To illustrate the potential gain, we also show in the left panel of figure~\ref{fig:eg_sum} a forecast scenario where $\beta$ is assumed to be known with 1\% precision at all redshifts. In this idealised case, percent-level precision on $E_G$ becomes achievable across a wide redshift range.

\subsection{Constraints on modified gravity}
\label{subsec:results_MG}
\begin{figure}[t!]
    \centering
    \includegraphics[width=0.48\textwidth]{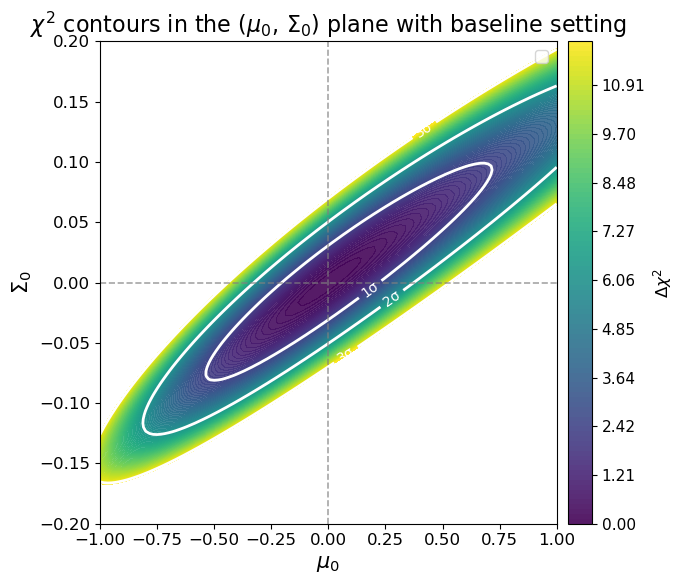}
    \hfill
    \includegraphics[width=0.48\textwidth]{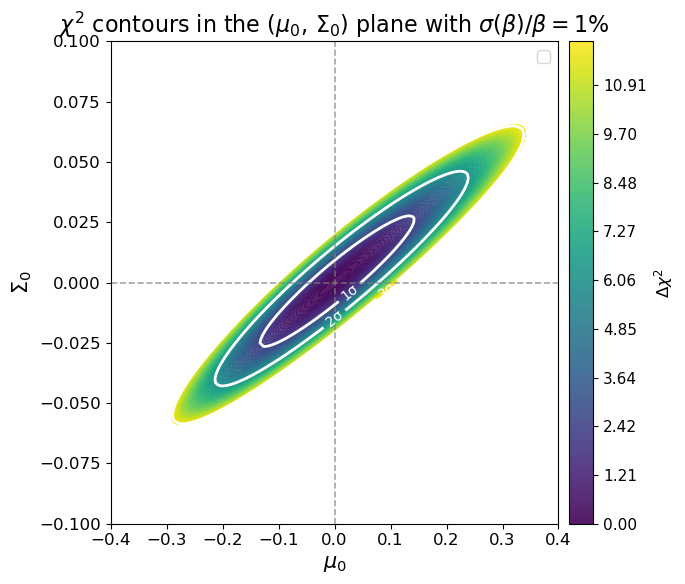}
    
    \caption{
    $\chi^2$ confidence contours in the $(\mu_0,\Sigma_0)$ plane,
    derived from the forecasted $E_G$ measurements.
    The contours represent the $68\%$ and $95\%$ joint confidence regions.
    \textit{Left:} Baseline result. $
    \mu_0 = 0^{+0.483}_{-0.324}, 
    \Sigma_0 = 0^{+0.064}_{-0.052}.
    $
    \textit{Right:} The case where $\sigma(\beta)/\beta$ is set to $1\%$.
    $
    \mu_0 = 0^{+0.0908}_{-0.0921}, 
    \Sigma_0 = 0^{+0.0177}_{-0.0184}.
    $
    }
    \label{fig:mu_sigma}
\end{figure}

This redshift-dependent summary of $E_G$ predictions highlights the overall constraining power of CSST across $0 < z < 1.5$, and provides a convenient tomographic view of the $E_G$ statistic for testing modified-gravity scenarios. To translate the forecasted $E_G$ measurements into anticipated constraints on modified gravity, we evaluate the likelihood of the phenomenological parameters $(\mu_0,\Sigma_0)$ entering  the $\mu-\Sigma$ parametrization described in section~\ref{subsec:mu_simga_param}. Figure~\ref{fig:mu_sigma} shows the resulting $\chi^2$ contours in the $(\mu_0,\Sigma_0)$ plane. 


We first note a strong degeneracy between $\mu_0$ and $\Sigma_0$, as expected from their respective roles in the denominator and numerator of equation~(\ref{eq:EG_muSigma}). The forecast constraints for the CSST baseline configuration are shown in the left panel of figure~\ref{fig:mu_sigma}. While $\mu_0$ can be constrained at the level of $30-50\%$, the precision on $\Sigma_0$ is significantly higher, reaching $5\%$. This indicates that $E_G$ is more sensitive to $\Sigma_0$ than to $\mu_0$, likely because the dependence on $\Sigma_0$ in the numerator of equation~(\ref{eq:EG_muSigma}) is direct and linear, whereas the dependence on $\mu_0$ in the denominator is indirect and nonlinear.  These baseline forecasts suggest that CSST will deliver competitive constraints on modified gravity parameters~\citep{Alam2021,ZhaoC2024,Frusciante2025,Leonard2025,Viglione2025}.


This potential is further underscored by an idealized scenario with $\sigma(\beta)/\beta = 1\%$ (right panel of figure~\ref{fig:mu_sigma}), which yields 5 times tighter constraints on both parameters. In particular, by leveraging improved $\beta$ measurements---anticipated at the percent level---from spectroscopic surveys like MUST~\citep{ZhaoC2024} and WST~\citep{Mainieri2024}, it becomes feasible to achieve percent-level constraints on $\Sigma_0$, the parameter associated with the effective gravitational constant of the Weyl potential. Such a measurement would provide a stringent direct test of the strength of gravity on cosmological scales, offering a critical benchmark to distinguish GR from modified gravity models.

\begin{figure}[t!]
    \centering
    \includegraphics[width=0.75\textwidth]{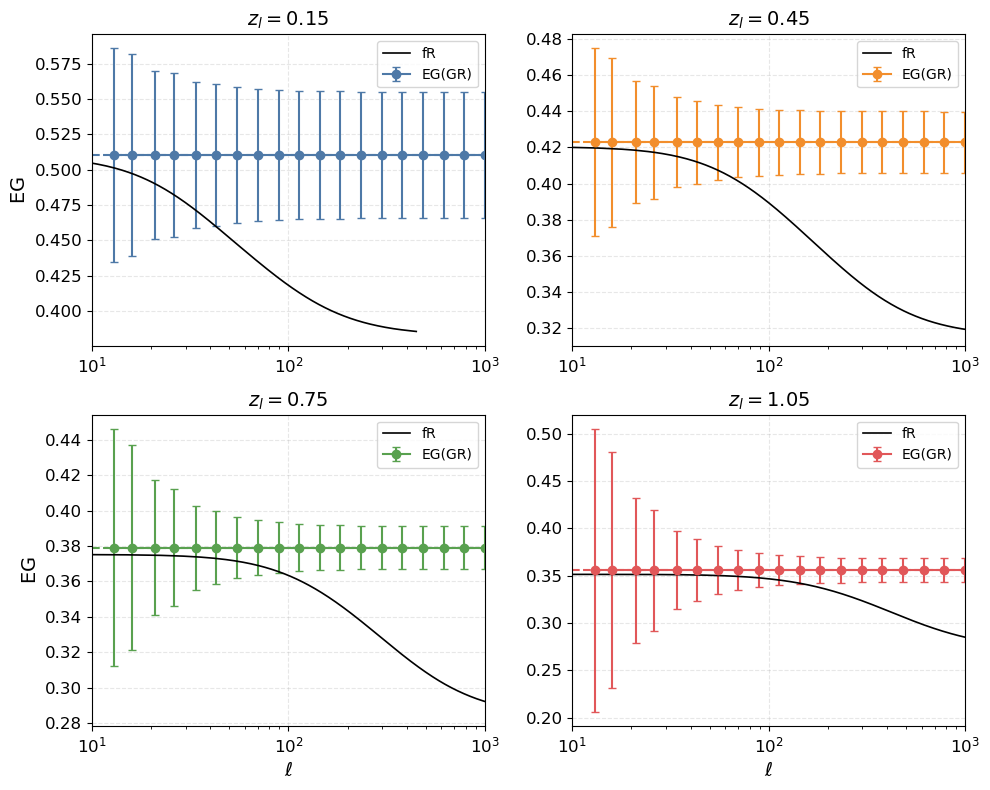}
    \caption{The behavior of $E_G$ in $f(R)$ gravity ($|f_{R0}| = 10^{-5}$). The deviation from GR becomes significant at large multipoles ($\ell$), while $E_G$ approaches the GR prediction at small $\ell$.}
    \label{fig:FR}
\end{figure}

\begin{figure}[t!]
    \centering
    \includegraphics[width=0.75\textwidth]{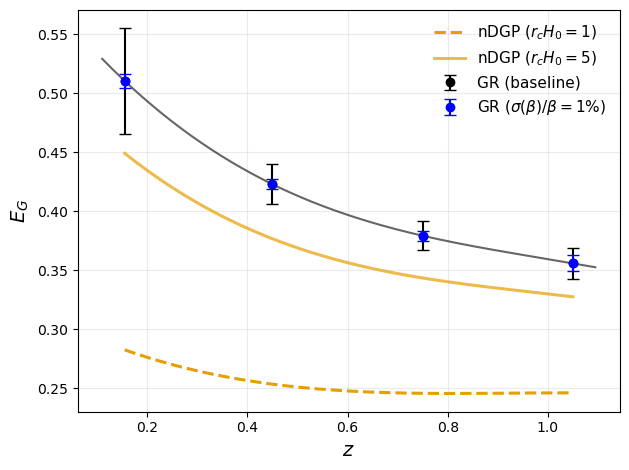}
    \caption{Redshift evolution of $E_G$ in the nDGP model. The curves correspond to $r_c H_0 = 1$  and $r_c H_0 = 5$, together with the GR prediction. The case $r_c H_0 = 1$ shows a more pronounced deviation from GR, while $r_c H_0 = 5$ remains closer to the GR limit.}
    \label{fig:nDGP}
\end{figure}
As shown in figure~\ref{fig:FR}, the $E_G$ statistic in $f(R)$ gravity exhibits a clear scale dependence in harmonic space. At large multipoles (small physical scales), $E_G^{f(R)}$ is significantly suppressed relative to the General Relativity (GR) prediction, whereas at low $\ell$ it gradually converges to the scale-independent GR value. This behavior reflects the scale-dependent modification of gravity in $f(R)$ models, where deviations become more pronounced on smaller scales. In contrast, $E_G$ is intrinsically scale-independent in GR, making its angular dependence a direct probe of modified gravity. The CSST forecasts shown in figure~\ref{fig:FR} demonstrate a strong capability to discriminate between GR and $f(R)$ gravity (with $|f_{R0}| = 10^{-5}$) at small scales.


Figure~\ref{fig:nDGP} compares the redshift evolution of the CSST $E_G$ forecasts with the predictions of the nDGP model, which yields a scale-independent $E_G$ with a different amplitude than the GR case. The nDGP model considered here predicts systematically lower $E_G$ values than GR at all redshifts, with deviations exceeding the $1\sigma$ uncertainty level, making them detectable given the precision expected from CSST.



\section{Conclusion}
\label{sec:conclusion}

In this study, we have systematically forecast the potential of the CSST survey to measure the $E_G$ statistic, a powerful multi-tracer test of gravitational physics on cosmological scales. By adopting redshift distributions of realistic mock spectroscopic and photometric galaxy samples, we developed a comprehensive harmonic-space covariance framework that fully accounts for the cross-correlation between galaxy--galaxy lensing and galaxy clustering observables.

Our analysis yields several key quantitative outcomes:

\begin{itemize}
    \item CSST will achieve a few-percent level precision on $E_G$ over a wide range of angular scales, with constraints reaching $\sim 4\%$ at $\ell > 100$ across various lens redshift bins. By compressing information of various scales and source redshift bins, the predicted scale-independent $E_G$ measurements can reach $3\%-9\%$ at $0<z<1.2$ --- a factor of several to an order of magnitude improvement over current observations.

    \item Within the $\mu-\Sigma$ parametrization of modified gravity, CSST can constrain $\Sigma_0$---the parameter associated with the effective gravitational constant of the Weyl potential---to $\sim 5\%$, while $\mu_0$ is constrained at the $30-50\%$ level.

    \item The dominant source of uncertainty in $E_G$ originates from the RSD parameter $\beta$, whose precision limits the overall constraining power. If future spectroscopic surveys improve $\beta$ constraints to the $1\%$ level, $E_G$ can be also measured at the percent-level over a wide redshift range, and both $\mu_0$ and $\Sigma_0$ could be constrained about five times more tightly, with $\Sigma_0$ approaching percent-level precision.

    \item For specific modified gravity models, the $f(R)$ scenario exhibits a pronounced scale dependence in $E_G(\ell,z)$, with a clear suppression relative to GR at high multipoles, while converging to GR on large scales. Such scale-dependent features are well within the sensitivity of CSST.

    \item In contrast, the nDGP model primarily induces a nearly scale-independent shift in $E_G$, yielding systematically lower values than GR across all redshifts, with deviations exceeding the $1\sigma$ level and thus remaining distinguishable with CSST measurements.
\end{itemize}

These results confirm that CSST will serve as a highly competitive Stage-IV facility for tomographic tests of gravity. The $E_G$ statistic, benefiting from the inherent cancellation of cosmic variance and independence from galaxy bias and $\sigma_8$, offers a robust and model-independent diagnostic for deviations from GR. Our forecasts further highlight the critical synergy between photometric weak lensing surveys and high-precision spectroscopic campaigns: improvements in $\beta$ determination directly and substantially enhance the $E_G$ constraining power.

Looking ahead, the methodology and baseline predictions presented here will provide a ready framework for the cosmological interpretation of real CSST data. Subsequent analyses will need to incorporate more realistic treatments of photometric redshift errors, intrinsic alignments, lensing magnification and higher-order systematics~\citep{Frusciante2025}. In combination with contemporaneous surveys such as Euclid~\citep{Euclid}, LSST~\citep{LSST2012}, DESI~\citep{DESI2016}, MUST~\citep{ZhaoC2024} and WST~\citep{Mainieri2024}, CSST is poised to deliver transformative constraints on modified gravity models and advance our understanding of cosmic acceleration.

\begin{acknowledgements}
We thank Zhejie Ding and Yan Gong for helpful discussions. 
Y.Z. acknowledges the support from the National SKA Program of China (2025SKA0150104) 
and the National Natural Science Foundation of China (NSFC) through grant 12203107.
\end{acknowledgements}

\bibliography{mybib}     
\end{document}